\documentclass[sigconf, preprint]{acmart}

\usepackage{amsmath,amssymb,amsfonts}
\usepackage{graphicx}
\usepackage{textcomp}
\usepackage{xcolor}
\def\BibTeX{{\rm B\kern-.05em{\sc i\kern-.025em b}\kern-.08em
    T\kern-.1667em\lower.7ex\hbox{E}\kern-.125emX}}

\usepackage{comment}
\usepackage{graphicx}
\usepackage{grffile}
\usepackage{url}
\usepackage{enumerate}
\usepackage{xspace}
\usepackage{subcaption}
\usepackage{algorithm}
\usepackage{algorithmicx}
\usepackage{algpseudocode}
\usepackage{etoolbox}\AtBeginEnvironment{algorithmic}{\small}
\usepackage{booktabs} 
\usepackage{appendix}
\usepackage{makecell}

\newcounter{note}[section]

\newcommand{\notecolor}{blue}
\renewcommand{\thenote}{\thesection.\arabic{note}}

\newcommand{\todo}[1]{\refstepcounter{note}{\bf \textcolor{\notecolor}{$\ll$TODO~\thenote: {\sf #1}$\gg$}}}

\newcommand{\myparagraph}[1]{\vspace{0.5em}\noindent\textbf{#1:}}
\newcommand{\myparagraphi}[1]{\vspace{0.5em}\noindent\textit{#1:}}

\newcommand{\prob}[1]{\ensuremath{\mathbb{P}\left({#1}\right)}\xspace}
\newcommand{\successEvent}{\ensuremath{\mathsf{Success}}\xspace}

\newcommand{\SSet}{\mathcal{S}\xspace}
\newcommand{\Tag}{\mathcal{T}\xspace}
\newcommand{\variants}{\ensuremath{n}\xspace}
\newcommand{\variantIdx}{\ensuremath{i}\xspace}
\newcommand{\replicaIdx}[1]{\ensuremath{j_{#1}}\xspace}
\newcommand{\replicasPerVariant}{\ensuremath{m}\xspace}
\newcommand{\fractionAdvReqs}{\ensuremath{\alpha}\xspace}
\newcommand{\rejuvenatedPerReq}{\ensuremath{k}\xspace}
\newcommand{\rejuvenatedPerAdvReq}{\ensuremath{b}\xspace}
\newcommand{\compromised}[1]{\ensuremath{c_{#1}}\xspace}
\newcommand{\compromisedFraction}{\ensuremath{\gamma}\xspace}

\newcommand{\cleansedPerRequestRV}[1]{\ensuremath{X_{#1}}\xspace}
\newcommand{\hypergeometric}[3]{\ensuremath{\mathrm{hypergeometric}({#1},{#2},{#3})}\xspace}
\newcommand{\balancedBins}{\ensuremath{n}\xspace}
\newcommand{\balancedBinsSample}{\ensuremath{d}\xspace}
\newcommand{\kn}{\ensuremath{\frac{\rejuvenatedPerReq}{\variants}}\xspace}
\newcommand{\ka}{\ensuremath{\frac{\rejuvenatedPerReq}{\fractionAdvReqs}}\xspace}
\newcommand{\rollingSum}{\ensuremath{\mathrm{K}}\xspace}
\newcommand{\throughputPerVCPU}{\ensuremath{\tau}\xspace}

\newcommand{\system}{\variants-\replicasPerVariant-Variant System\xspace}
\newcommand{\systems}{\variants-\replicasPerVariant-Variant Systems\xspace}

\AtBeginDocument{%
  \providecommand\BibTeX{{%
    \normalfont B\kern-0.5em{\scshape i\kern-0.25em b}\kern-0.8em\TeX}}}

\setcopyright{none}
\copyrightyear{}
\acmYear{}
\acmDOI{}

\acmConference[]{}{}{}
\acmBooktitle{}
\acmPrice{}
\acmISBN{}

\renewcommand\footnotetextcopyrightpermission[1]{}

\begin{document}
\title{\systems: Adversarial-Resistant Software Rejuvenation for Cloud-Based Web Applications}

\author{Isaac Polinsky}
\affiliation{North Carolina State University}
\email{ipolins@ncsu.edu}
\author{Kyle Martin}
\affiliation{North Carolina State University}
\email{kdmarti2@ncsu.edu}
\author{William Enck}
\affiliation{North Carolina State University}
\email{whenck@ncsu.edu}
\author{Michael K. Reiter}
\affiliation{UNC-Chapel Hill}
\email{reiter@cs.unc.edu}

\begin{abstract}
  Web servers are a popular target for adversaries as they are publicly accessible and often vulnerable to compromise.
  Compromises can go unnoticed for months, if not years, and recovery often involves a complete system rebuild.
  In this paper, we propose \systems, an adversarial-resistant software
  rejuvenation framework for cloud-based web applications. We improve
  the state-of-the-art by introducing a variable \replicasPerVariant
  that provides a knob for administrators to tune an environment to
  balance resource usage, performance overhead, and security guarantees.
  Using \replicasPerVariant, security guarantees can be tuned for seconds,
  minutes, days, or complete resistance.
  We design and implement an \system prototype to protect a
  Mediawiki PHP application serving dynamic content from an
  external SQL persistent storage.
  Our performance evaluation shows a throughput reduction of 65\% for 108 seconds of resistance and 83\% for 12 days of resistance to sophisticated adversaries, given appropriate resource allocation.
  Furthermore, we use theoretical analysis and simulation to characterize the impact of system parameters on resilience to adversaries.
  Through these efforts, our work demonstrates how properties of cloud-based servers can enhance the integrity of Web servers.
\end{abstract}

\maketitle

\pagestyle{plain}

\excludecomment{techreport}
\includecomment{conference}

\section{Introduction}

Web servers consist of large code bases that are difficult to verify and frequently contain
exploitable vulnerabilities. Compromised servers can go unnoticed for months, if not
years~\cite{fireeye}. During this time, adversaries may maliciously modify persistent storage to
persist through reboots (e.g., rootkit) or host malicious content (e.g., watering hole attacks).
Once a compromised server is discovered, a time consuming process is performed to identify what
files were affected by the adversary and restore the server to a good state.

Periodically refreshing a server to a known good state 
can time bound compromise. This process is commonly known as \emph{software
rejuvenation}~\cite{has95}. In fact, cloud environments are particularly amenable to software
rejuvenation, because they frequently boot from read-only images and use external persistent
storage. However, traditional software rejuvenation cannot tolerate an adversarial threat model: if
a compromised server maliciously modifies persistent storage, then subsequent reads from persistent
storage may automatically re-compromise instances after refresh. Thus, a key challenge is to ensure
that persistent storage is not maliciously modified.

Software Diversity is a promising approach to prevent malicious changes to persistent storage in a
software rejuvenation setting. Byzantine Fault Tolerance (BFT)~\cite{lsp82, ack+10, cb99},
\variants-Variant Systems~\cite{cef+06}, and Multi-Variant Execution Environments
(MVEEs)~\cite{vcs16, sjg+09, bz06} run multiple, functionally identical, but internally diverse,
instances of software independently to detect abnormal behavior, such as faults or compromise. By
comparing writes to storage from each instance, software diversity can detect and prevent malicious
behavior then trigger software rejuvenation to reactively recover affected instances. Meanwhile,
periodic refreshing~\cite{sbc+10} can proactively recover undetected compromised instances. However,
these existing approaches have several limitations when applied to a concurrent Web server
environment. When \variants-Variant Systems and MVEEs detect a difference between the instances, the
entire system goes offline until refreshing is complete. Similarly, BFTs require that no more than
$f$ servers are malicious at any given moment, and so if $f$ servers are offline for refreshing, no
security guarantees can be made and new requests must wait for recovery to complete. Furthermore, no
existing system based on BFT, \variants-Variant System, or MVEE can defend against a powerful
adversary that discretely compromises servers, one-by-one, until sufficient instances are acquired
to modify persistent storage without detection.

In this paper, we propose \systems for enhancing the resilience of cloud-based servers. \systems
extend BFT and MVEE-based software rejuvenation systems by introducing the variable
\replicasPerVariant, which is a pool of active replicas for each of the \variants diverse variants.
The introduction of \replicasPerVariant provides several key properties. First, \replicasPerVariant
increases availability by allowing processing to continue while some replicas are offline for
refreshing. Second, \replicasPerVariant minimizes the need to create many diverse variants, while
increasing availability. Third, \replicasPerVariant simplifies the diverse computing architecture
when addressing highly concurrent workloads. Finally, \replicasPerVariant, combined with our
configurable refreshing algorithm, provides a knob that allows an administrator to balance resource
usage, performance overhead, and security guarantees. This knob can tune security for seconds,
minutes, days, or complete resistance.

We built a prototype of our \systems design for web servers in a cloud environment. To demonstrate
feasibility, we instrumented two web server stacks with a host agent: Apache on Linux and IIS on
Microsoft Windows. Further, we evaluated our system using the Linux prototype hosting Mediawiki, a
popular wiki application. Using Apache JMeter~\cite{jmeter} and varying resource configurations, we
show our prototype incurs a throughput reduction of 65\% for 108 ``seconds of resistance'' and 83\%
throughput reduction for 12 ``days of resistance'' when refreshing half of the servers used in every
HTTP request. Finally, we illustrate how an administrator can tune parameters to balance risk
tolerance with performance overhead and resource cost (e.g., the above calculations show the
overhead assuming 10\% of traffic is malicious).

We make the following contributions in this paper.
\begin{itemize}
  \item \textit{We enhance defense techniques that combine software rejuvenation and software
    diversity by increasing their availability.} \systems introduces the variable
    \replicasPerVariant and
    can increase the availability of BFT and MVEE systems while performing recovery actions to
    defend against a powerful adversary.

  \item \textit{We provide a theoretical security analysis of \systems.}
    The \replicasPerVariant configuration provides a knob that increases the availability of servers
    while accommodating for greater security configurations. By using the balanced allocation
    problem and matching simulations, we model the security impact of \variants,
    \replicasPerVariant, and the refreshing strategy. 

  \item \textit{We demonstrate the feasibility of \systems through empirical analysis.}
    Our prototype for a Mediawiki application with a remote SQL database
    has a 65\% reduction in throughput and a 83\% increase in latency for a 2-25-Variant environment
    and a 83\% reduction in throughput and a 111\% increase in latency for a 4-15-Variant
    environment.
\end{itemize}

The remainder of this paper proceeds as follows. Section~\ref{sec:relwork} discusses background and
related work. Section~\ref{sec:overview} overviews the concept of \systems. Section~\ref{sec:design}
describes our design. Sections~\ref{sec:seceval} and~\ref{sec:eval} evaluate the security and
performance of our prototype. Section~\ref{sec:discussion} discusses trade-offs and limitations.
Section~\ref{sec:conc} concludes.
 
\section{Related Work}
\label{sec:relwork}

Refreshing hardware or software systems to mitigate flaws before their
manifestation is termed \emph{software rejuvenation}~\cite{has95} in the
fields of fault tolerance and reliability. Machida et al.~\cite{mkt10}
propose using software rejuvenation to enhance the availability of
virtual machines and virtual machine monitors.  Similarly, Rezaei et
al.~\cite{rs10} and Thein et al.~\cite{tcp+08} argue to
periodically refresh systems in a virtual environment to remove
inconsistent states.

In general, software rejuvenation alone cannot be used
as a defense mechanism against powerful adversaries. For example,
CRIU-MR~\cite{wep18} uses software rejuvenation to quickly remove
malware from a system, but since it is unable to determine which writes
to storage are malicious, their conservative solution results in benign writes
being reverted when the malware is removed. Therefore, software
rejuvenation must be combined with other techniques to detect and
prevent malicious actions and then reactively recover. Additionally,
these techniques can use proactive recovery to refresh a compromised
system before it is detected. Sousa et al.~\cite{sbc+10} enhances such
proactive recovery techniques with an approach aiming to keep a
minimal number of systems online to ensure the correct operation of a
firewall. However, their work is tied to wall clock time and is not
suitable for high request Web servers. Brand\~{a}o et al.~\cite{bb11}
provide an analysis of intrusion-tolerant systems built on software
rejuvenation and state that intrusion is inevitable for systems that do not
recover on each request. In this work, we report the same finding, but
define a model that fits our architecture.

Software diversity can be used to detect and prevent malicious behaviors.
Approaches such as MVEEs and Intrusion Tolerant Replication based on BFT 
detect abnormal behavior by comparing outputs (system calls, server
responses, etc.) from diverse systems. If a divergence between outputs is
detected, then these systems can prevent the action. \systems is
based on \variants-Variant Systems~\cite{cef+06} and similar MVEEs that use
\variants diversity to process requests independently. This use of
diverse systems was first proposed as \variants-Version
Programming~\cite{abc94, avi85, ca78} to detect software faults.

Many MVEEs have been proposed: Chun et al.~\cite{cms08} propose an
architecture using virtualization on a  single physical host. Salamat
et al.~\cite{sjw+11} defend against code injection attacks with a
user-space architecture. GHUMVEE~\cite{vsb+12}, ReMon~\cite{vcv+16},
Orchestra~\cite{bcl07}, and Volckaert et al.~\cite{vcs+17} all create
architectures for complex threaded-processes. VARAN~\cite{hc15} seeks to
increase performance by avoiding costly system call lockstepping. 
DREME~\cite{bee13} defends against SQL injection attacks by using redundant
database variants and diverse processes. HACQIT~\cite{rjl+02} uses server
diversity (IIS and Apache) to mediate storage accesses from vulnerable web
applications and introduces replay attack prevention using blacklists.
Finally, Gao et al.~\cite{gks09} and
STILO~\cite{xyr+15} use probabilistic anomaly detection over system calls to
identify misbehaving variants. In addition, Intrusion Tolerant Replication
techniques, specifically those based on BFT, are another way of
employing diversity to prevent malicious behavior. Spire~\cite{bta+18},
Steward~\cite{add+10}, and Base~\cite{rcl01} all propose techniques to
tolerate a compromised host until it is recovered. However, the main
drawback of BFT-based approaches is the low number of compromised hosts
BFT protocols can tolerate until an undetected adversary can compromise the
entire system.

We do not aim to replace prior MVEE and BFT-based Intrusion Tolerant
systems but rather to enhance their architectures to increase
availability. In previous works, the entire system goes down when
performing recovery actions, which is unacceptable for web servers. In
\systems there are \replicasPerVariant replicas of each variant that can
continue processing when a replica goes offline. Further, we prioritize
securing dynamic web applications. Prior works have considered web servers
hosting static files and SCADA systems, but to the best of our knowledge we
are the first to propose an architecture for a dynamic web server. There are
two notable differences between dynamic servers and the servers targeted in
prior works: (1) generating dynamic content requires more I/O
operations, resulting in more overhead and (2) comparing database writes
from replicas to detect malicious writes requires that any
non-deterministic fields in those writes be reconciled.  Finally, we note
that creating diversity is not a contribution of this work. Weatherwax
et al.~\cite{wkn09} define guidelines for implementing \variants-Variant
Systems that defend against particular attacks. Further, works that
describe the creation of diverse replicas~\cite{bz06, clh+15, sgf08,
hjc+17} are complementary to this work and can be used by our architecture.

\section{Overview}
\label{sec:overview}

This work seeks to provide resiliency for cloud-based servers by periodically refreshing replicas
from read-only images to automatically remove persistent threats (e.g., rootkits). A na\"ive
application of software rejuvenation is not resilient to sophisticated adversaries that craft and
commit exploits into external persistent storage, with the goal of automatically re-compromising
refreshed replicas. 
This threat model presents the following research challenges.

\begin{itemize}

  \item \textit{Identifying malicious writes to external persistent storage.}
    Preventing malicious writes to persistent storage prevents refreshed replicas from being
    automatically re-compromised.

  \item \textit{Providing a tunable model for balancing security and resource overhead.}
    Frequently refreshing replicas incurs resource overheads for service providers. Providers should
    be given guidance on how to provision resources to meet their risk model.

  \item \textit{Providing a high availability environment.}
    Solutions should provide a highly available Web environment in the face of compromise and limit
    the impact on throughput and latency.

\end{itemize}

We address the first challenge by adopting the concept of \variants-Variant Systems~\cite{cef+06}
and extending it to a cloud environment. Consider a Web server that uses an external relational
database. Typically, each HTTP request to the Web server will result in one or more SQL queries to
the relational database. In our model, each HTTP request from a client is duplicated and sent to
$\variants$ different server variants. We then compare each of the resulting SQL queries to ensure
that all $\variants$ variants match, modulo non-deterministic fields such as timestamps. If any one
SQL query differs from the query produced by the $\variants-1$ other variants, the query fails
(i.e., preventing any writes) and the corresponding replicas for all $\variants$ variants are
refreshed from the good, read-only image.

We address (Section~\ref{sec:eval_cost}) the second challenge by combining our performance
evaluation (Section~\ref{sec:refreshing-eval}) with the theoretical analysis
(Section~\ref{sec:seceval}). Finally we address the third challenge by introducing a new variable,
$\replicasPerVariant$, to the traditional \variants-Variant architecture described above. With this
variable, a deployment can add additional servers which all concurrently handle requests. Further,
\replicasPerVariant allows the environment to continue handling requests if any given server goes
offline (e.g., is refreshed as described above). Effectively, \replicasPerVariant allows us to
provide high availability and throughput that is not achievable in traditional \variants-Variant
Systems.

\begin{figure}[t]
  \centering
  \includegraphics[width=1.00\columnwidth]{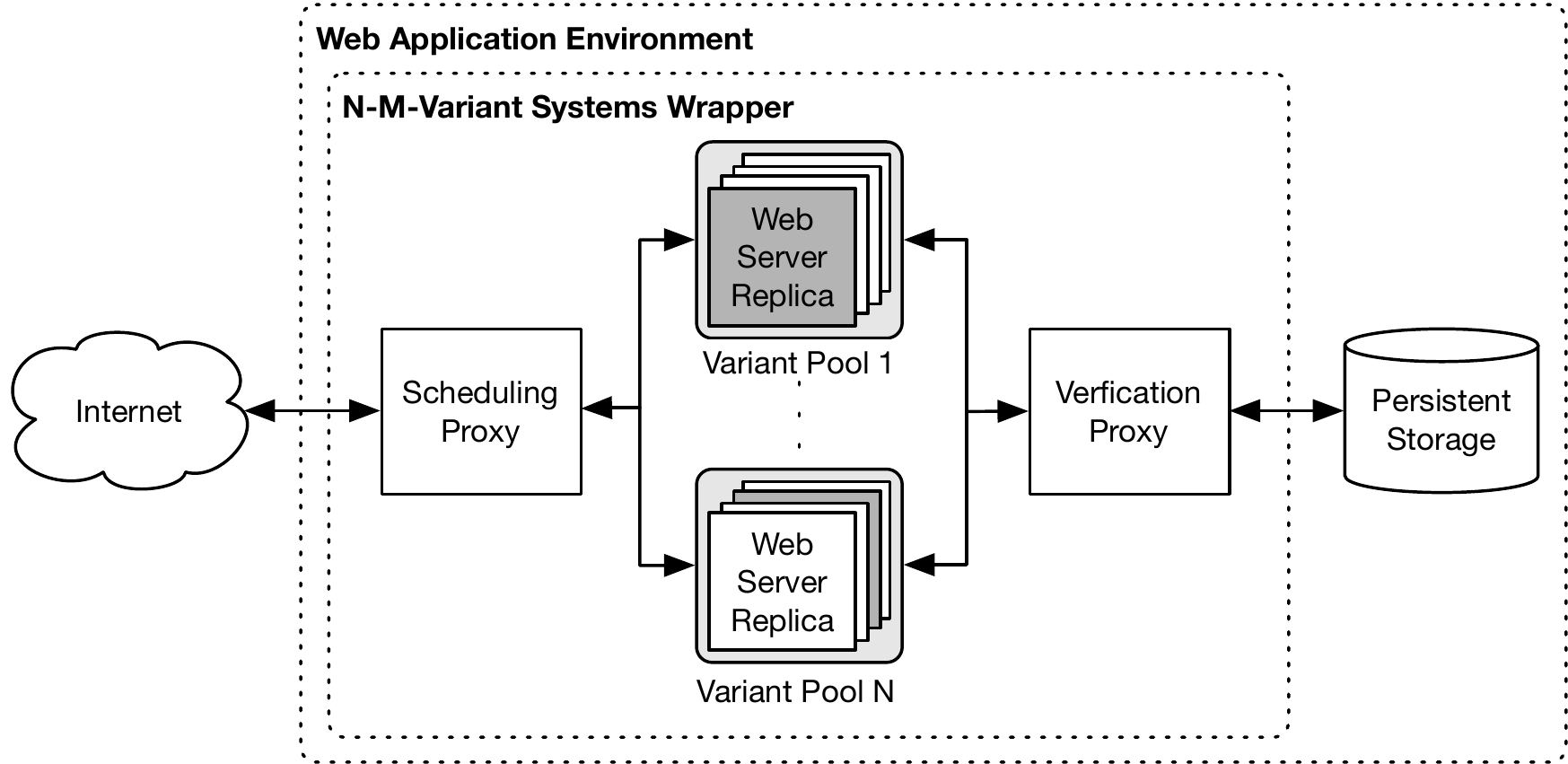}
  \caption{\systems for a Web application.}
  \label{fig:overview}
\end{figure}

\begin{conference}
Figure~\ref{fig:overview} depicts an \system for a Web environment using an external relational
database for persistent storage. HTTP requests from Web clients terminate at a Scheduling Proxy. The
Scheduling Proxy selects a serving set of replicas, which is one of the \replicasPerVariant replicas
from each of the \variants variant pools. The HTTP request is duplicated and processed by each
serving set replica. All SQL queries from the replicas are captured by the Verification Proxy. If
the SQL queries differ (modulo non-deterministic fields), a malicious write is detected, and all
replicas in the serving set are immediately refreshed. However, if the SQL queries are the same, the
Verification Proxy sends one SQL query to the persistent storage. The Verification Proxy then
duplicates the SQL response to each replica. When the replicas finish generating the Web page, the
HTTP responses are returned to the Scheduling Proxy, which returns one HTTP response to the Web
client. Finally, replicas are periodically refreshed (Section~\ref{sec:replica_refresh}) to mitigate
an adversary who attempts to gradually compromise replicas in an effort to be assigned a serving set
with an already compromised replica in each of the \variants variant pools.
\end{conference}

\myparagraph{Threat Model and Assumptions}
The goal of the adversary is to compromise and \emph{maintain} privileged access to a vulnerable
cloud-based Web server. Once compromised, a replica may serve malicious content to clients or
attempt malicious modifications to persistent storage. The goal of this work is to minimize the
duration of a compromised server by preventing malicious changes to persistent storage. We permit a
time-bound period where Web clients are served malicious pages. We also do not attempt to prevent
exfiltration of data present at the time of compromise.
The adversary is assumed to be external to the cloud environment and must craft unprivileged HTTP
requests to compromise the server and perform arbitrary code execution on the Web server, including
running code in kernel, but not the hypervisor. We assume a single malicious HTTP request cannot
simultaneously compromise replicas of different variants. This means vulnerabilities that effect the
same application on diversified hosts (i.e., SQL injection) cannot be stopped; however leveraging
previous MVEE works we can defend Memory Corruption vulnerabilities or other vulnerabilities that
lead to Remote Code Execution on the Web server. This work assumes only the base static Web server
image is trusted and the running Web server instance can be compromised during the processing of any
request. The trusted computing base includes the scheduling and verification proxies, the external
persistent storage, the cloud infrastructure (i.e., management software, hypervisors, hardware,
personnel), and Web application administrators.

\section{Design}
\label{sec:design}

In order to describe concrete design decisions for \systems, we describe our design with respect to
a Web application environment using a relational database for external persistent storage. However,
the high-level conceptual approach is more generic.

\subsection{Scheduling Proxy}

Our \system-based Web application operates on the granularity of HTTP requests. The Scheduling
Proxy, $P_s$, is the only interface between Web clients and the \system. When $P_s$ receives an HTTP
request, $R_h$, it performs the following tasks: (a)~it selects a serving set, $\SSet(R_h)$, for
$R_h$, consisting of one server replica from each variant pool, i.e., $\{S_{1,\replicaIdx{1}},
S_{2,\replicaIdx{2}}, \ldots, S_{\variants,\replicaIdx{\variants}}\}$; (b)~it duplicates $R_h$,
sending a copy to each server replica $S \in \SSet(R_h)$; and (c)~it tags each duplicate of $R_h$
with an unforgeable request identifier $\Tag(R_h)$ that binds the IP addresses of the server
replicas to the identifier. We begin with serving set selection.

\subsubsection{Serving Set Selection}

The serving set, $\SSet(R_h)$, consists of one server replica from each variant pool. We identify
each server replica as $S_{\variantIdx,\replicaIdx{}}$, where \variantIdx is the variant pool index,
and \replicaIdx{} is the index of the replica within variant pool \variantIdx. 
\begin{conference}
Given an \variants and \replicasPerVariant configuration, it is possible to pre-compute all possible
serving sets for faster selection. In general, an \system has $\replicasPerVariant^\variants$ unique
serving sets.
\end{conference}

Our threat model assumes that an adversary cannot simultaneously exploit all variants with a single
request. Therefore, if each server replica, $S_{\variantIdx,\replicaIdx{}}$, is refreshed after
servicing a \emph{single} request, then it is not possible for the adversary to incrementally
compromise server replicas and opportunistically wait until a request is served by
\variants compromised server replicas, one in each pool. Unfortunately, only using a server replica
for a single request requires refreshing on average \variants server replicas per HTTP request,
which is cost prohibitive in most scenarios. A larger \replicasPerVariant simply provides a larger
buffer to handle bursts of requests.

Practical deployments must re-use a server replica, $S_{\variantIdx,\replicaIdx{}}$, for multiple
requests. Section~\ref{sec:seceval} further explores the security implications of re-use. Our
current implementation selects a serving set by randomly selecting one replica from each variant
pool. If the selected replica index, $S_{\variantIdx,\replicaIdx{}}$, is marked for refresh but not
yet refreshed, our algorithm randomly selects another until an available replica is found. If a
serving set cannot be selected, because at least one variant pool has no available replica (i.e.,
all are being refreshed), our system returns HTTP error code 503, Service Unavailable.

\subsubsection{Request Processing}
\label{sec:request}
\label{sec:requestidentification}

The Scheduling Proxy, $P_s$, intercepts the HTTP request, $R_h$, using a HTTP proxy. That is, the
Web client's HTTP connection is terminated at $P_s$, and $P_s$ initiates a new HTTP connection to
each replica $S \in \SSet(R_h)$. HTTP proxies are commonly used for load balancing. However, unlike
load balancing, $P_s$ must duplicate the HTTP payload into multiple requests. When the replicas
return the HTTP response, the HTTP proxy only returns one HTTP response to the client. $P_s$
currently does not compare the HTTP responses from the replicas, as our threat model only considers
the integrity of the external persistent storage and tolerates some period of malicious responses to
Web clients.

$P_s$ tags each connection to $S \in \SSet(R_h)$ with an unforgeable request identifier. This tag,
$\Tag(R_h)$, is read by the server replica host agent and propagated to the Verification Proxy,
$P_v$, as described in Section~\ref{sec:variant_pools}. $P_v$ needs $\Tag(R_h)$ to determine which
SQL queries belong to $R_h$. The unforgeable property is needed to meet our threat model requirement
of the server replica remaining untrusted. A compromised server may attempt two attacks: (1) Mimicry
attacks where an adversary attempts to send requests to persistent storage with a falsified
identifier, and (2) replay attacks where an adversary tries reusing an existing tag. The unforgeable
identifier created at $P_s$ prevents mimicry attacks and also makes replay attacks extremely
unlikely, as discussed below and in Section~\ref{sec:verification-packet}.

\begin{conference}
Communication of the request identifier from $P_s$ to each server replica is done by inserting
$\Tag(R_h)$ as an IP Option in the IP Header of the TCP SYN packet.
\end{conference}
Specifically, we use the timestamp option field, which supports up to 40 bytes, containing 36 bytes
of usable storage when accounting for the two bytes used for the IP header format and two bytes for
the timestamp option declaration.
The tag, $\Tag(R_h)$, consists of a 32-bit (4-byte) request identifier and a 256-bit (32-byte) HMAC.
The 32-bit request identifier, $ID(R_h)$, is a counter that increases with each HTTP request
received by $P_s$. The HMAC computes a hash over $ID(R_h)$ concatenated with the IP addresses of
each $S \in \SSet(R_h)$. The symmetric key $k$ is only shared by the $P_s$ and $P_v$. Therefore,
\[ \Tag(R_h) = [ ID(R_h)~|~HMAC(k,[ID(R_h)~|~\mathrm{IP}_1~|~\cdots~|~\mathrm{IP}_{\variants}]) ] \]
for each $\mathrm{IP}_{\variantIdx}$ corresponding to $S_{\variantIdx} \in \SSet(R_h)$. This tag is
a fixed size by design and the IP addresses in the serving set do not need to be explicitly listed
in the tag as they can be derived at $P_v$.  $P_v$ verifies the tag by collecting the IP addresses
of the server replicas declaring $ID(R_h)$ and recomputing the hash using the shared key.

Note that the 32-bit request identifier will roll-over every four billion HTTP requests. While the
adversary may attempt to exploit request identifier roll-over to replay a tag, the tag is only valid
if the request is served to server replicas with the exact IP addresses as the original request.
\begin{conference}
This can be further mitigated by implementing a validity window that rejects tags received with
identifiers outside of the current valid range.
\end{conference}

\subsubsection{Replica Refreshing}
\label{sec:replica_refresh}

\begin{conference}
The final responsibility of the scheduling proxy is coordinating the refreshing of replicas. 
\end{conference}
Replica refreshing is caused by a) malicious activity detected by the verification proxy, and b)
periodic refreshing of replicas. Since replicas can simultaneously serve multiple HTTP requests, the
design must decide whether the decision to refresh triggers immediate or delayed termination of the
replica VM. An immediate termination will cause collateral effects to other HTTP requests, impacting
the comparison at the verification proxy. When the verification proxy detects malicious activity,
our design uses immediate termination.
However, for periodic refreshes, the replica is marked for refresh but not terminated until all
current HTTP requests are processed. When marked for refresh, a replica will not be served new HTTP
requests. A reasonable timeout (e.g., 10 seconds) can also prevent adversaries from holding onto
replicas for a long period of time.

For periodic refreshing, our current design has the system administrator select $\rejuvenatedPerReq
> 0$ replicas to be refreshed after each HTTP request is serviced by the system. Since
$\rejuvenatedPerReq$ may be a fraction, the remainder is always carried forward to the next HTTP
request. Thus on each HTTP request, we add $\rejuvenatedPerReq$ to a rolling sum, $\rollingSum$. If
$\rollingSum < 1$, no replicas are refreshed. However if $\rollingSum \geq 1$, $\lfloor \rollingSum
\rfloor$ replicas are chosen across the entire set of $\variants\replicasPerVariant$ replicas to be
refreshed and the remainder is carried forward to the next request. Variant pools do not play a role
in the selection of replicas to be refreshed. If the randomly selected replica is already marked for
refresh or has not served any request since its last refresh, another replica is chosen.

\begin{conference}
  This refreshing strategy maps well to the theoretically grounded security evaluation presented in
  Section~\ref{sec:seceval}. Further, by tying refresh rate to the request rate we can defend
  against large bursts of malicious requests between a set refresh interval; however this opens up a
  potential avenue for a denial-of-service attack. Section~\ref{sec:discussion} addresses this issue
  in more detail.
\end{conference}

\subsection{Variant Pools}
\label{sec:variant_pools}

Each variant pool contains \replicasPerVariant server replicas. \systems assumes a single malicious
HTTP request cannot simultaneously exploit multiple variants. We assume variants are created using
existing techniques discussed in Section~\ref{sec:relwork} (e.g., automated software
diversification~\cite{hjc+17}) and do not detail the creation of variants. We chose virtual machines
over containers due to their stronger isolation.
\begin{conference}
We now describe the operation of the untrusted Host Agent. 
\end{conference}

\myparagraph{Untrusted Host Agent}
Each server replica runs an untrusted host agent that propagates the unforgeable request identifier,
$\Tag(R_h)$, from the IP options field in the received HTTP request to all outbound connections
created by the servicing process or thread. We place this logic within the server replica to
simplify our prototype implementation. While virtual machine introspection from the trusted
hypervisor is possible, significantly more effort is needed to correlate the inbound and outbound
connections. Furthermore, the unforgeable tag eliminates the need to trust the host agent. If a
malicious host agent does not include a verifiable tag, the Verification Proxy, $P_v$, will drop the
request.

For each HTTP request, $R_h$, the host agent must: (1) extract $\Tag(R_h)$ from the IP options in
the IP header of the TCP SYN packet for the request, (2) identify the process or thread identifier,
$PID_a$, processing $R_h$, (3) identify the process or thread identifier, $PID_b$, making the
corresponding TCP connection for SQL queries, and (4) insert $\Tag(R_h)$ in the IP options of the IP
header of the TCP SYN packet of all outgoing connections from $PID_b$. Note that $PID_a$ may equal
$PID_b$, or $PID_b$ may be a child (or descendant) of $PID_a$.

\myparagraphi{Linux Host Agent}
Our Linux host agent is a user space Python program that uses the Netfilter kernel interface to read
and insert request identifier tags in network connections. The host agent intercepts all outbound
database traffic with the TCP SYN flag set. It then determines which PID owns the socket for the
outbound request and traverses the PID's parents until it finds the PID of the process handling the
inbound request. Using this information, the host agent can determine the appropriate request
identifier for the outbound request and inserts it into the IP header. 

\myparagraphi{Windows Host Agent}
Our Windows host agent is a kernel mode driver, written in C, that uses the Windows Filtering
Platform (WFP) to read and insert request identifier tags in network connections. The host agent
also relies on an IIS HTTP module, written in C\#, and a DLL, written in C, injected into the IIS
process to record which PHP process is spawned to handle each inbound request. Similar to the Linux
host agent, the kernel driver, IIS module, and DLL allow the host agent to correlate inbound
connections to outbound connections and propagate tags to outbound request IP headers.

\subsection{Verification Proxy}
\label{sec:verification}

The Verification Proxy, $P_v$, prevents malicious writes to the external persistent storage. To
identify malicious writes, $P_v$ uses unanimous voting for writes from all \variants variants. That
is, if any variant diverges, the write is denied. Furthermore, $P_v$ does not attempt to determine
which variant is malicious; all server replicas $S \in \SSet(R_h)$ are marked for refresh.
This design is security conservative, as $\variants-1$ variants in the serving set may have been
compromised.

To perform voting, $P_v$ must be able to compare the write requests from the \variants variants in
$\SSet(R_h)$. To simplify this comparison, we assume that each variant is running the same
application, and produces nearly identical SQL queries to write to external persistent storage.
While the SQL queries may not be identical (e.g., timestamp fields), it is reasonable to identify
non-deterministic fields. Next we discuss the primary tasks of $P_v$: (1) determining which queries
to compare and (2) performing the comparison.

\subsubsection{Packet Processing}
\label{sec:verification-packet}

The Verification Proxy, $P_v$, uses a TCP proxy to intercept network connections destined for the
external SQL server. When replica variant $i$ makes an SQL request, $R_s$, to $P_v$, the TCP proxy
inspects the IP options of the IP header of the TCP SYN packet. From here, it extracts the
unforgeable request identifier tag, $\Tag(R_h)$. Recall that $\Tag(R_h)$ contains the plaintext
32-bit request identifier, $ID(R_h)$, but not the list of IP addresses needed to compute the HMAC.
Therefore, $P_v$ maintains a queue for each $ID(R_h)$, storing the SQL query, source IP address, and
$\Tag(R_h)$ for each received $R_s$. When $P_v$ receives an SQL request, $R_s$, from each of the
\variants servers, it computes the HMAC and verifies that $\Tag(R_h)$ was not forged on any received
$R_s$ associated with $ID(R_h)$. If the tags verify, $P_v$ proceeds to verify the SQL query, as
described in Section~\ref{sec:verification-sql}. If the SQL query verification also succeeds, a
single SQL query is sent to the SQL server. The SQL response is duplicated and returned to each of
the replicas $S \in \SSet(R_h)$ in the queue.

If either the tag verification or the SQL query verification fails, $P_v$ assumes that one or more
of the server replicas is compromised. If this occurs, the SQL query is not sent to the SQL server.
Further, all server replicas $S \in \SSet(R_h)$ are marked for immediate refresh.

If $P_v$ does not receive all \variants SQL queries within a predefined timeout period, the queue
for $ID(R_h)$ is deleted and the SQL query is not sent to the SQL server. However, in this case, the
server replicas are not refreshed. Not receiving all \variants SQL queries before the timeout may
result from slow processing or network connectivity. Marking all server replicas $S \in \SSet(R_h)$
for refresh would further reduce available computation, 
causing a significant collateral effect for replicas simultaneously servicing multiple HTTP
requests.

\begin{conference}
  Note that OpenStack prevents guest machines from spoofing their IP address by default. Therefore,
  the adversary cannot spoof its IP address to fool the verification proxy. We assume other cloud
  environments have similar anti-spoofing defenses in place.
\end{conference}

Finally, each HTTP request, $R_h$, may result in multiple SQL requests, $R_{s1} \dots R_{si}$, as
the Web application code queries the database for various information. Our current implementation
assumes that the SQL queries from each of the \variants variants are received in the same order, as
was the case for the Mediawiki application used in our evaluation (Section~\ref{sec:eval}).
Theoretically, the order could be recovered by inspecting the queries themselves; however, the query
order may or may not have a logical impact on the Web application. Therefore, we leave out-of-order
query processing as a topic for future work.

\subsubsection{Query Matching}
\label{sec:verification-sql}

The Verification Proxy, $P_v$, prevents malicious writes to the external SQL server by comparing SQL
queries from each of the \variants variants. Once the request identifier tag is validated
(Section~\ref{sec:verification-packet}), the SQL query string is extracted from the request, $R_s$.
A na\"ive approach to query matching is simple string comparison. In practice, SQL queries contain
non-deterministic fields, such as timestamps, which are generated by the server replica. Even with
time synchronization, it is unlikely that all \variants server replicas will generate the same
timestamp.

To account for non-deterministic fields, $P_v$ extracts an abstract syntax tree (AST) from the query.
Our implementation uses a PostgreSQL parsing engine~\cite{libpg-query}, which limits our prototype
to Web applications compatible with PostgreSQL; however, it is feasible to integrate our system with
other databases. Once the AST is extracted for a given SQL query, $P_v$ recursively traverses the
tree looking for known non-deterministic fields. When a non-deterministic field is found, $P_v$
replaces the value with a constant value. Once the traversal is complete, $P_v$ collapses the
modified SQL query back into a string and uses string comparisons to perform the final match.

This algorithm requires non-deterministic fields to be known before deployment. Our prototype uses a
policy configuration file to define the non-deterministic fields for each table in the database
schema. Currently, we leave this policy definition as a manual process, requiring an administrator
or developer to identify the non-deterministic fields. Fortunately, defining this policy is a
one-time effort per Web application. For our evaluation we created our list of non-deterministic
fields by allowing queries with fields that did not match to proceed and then writing these fields to
a log. We then manually analyzed the log file to define the policy.

Our threat model only considers the integrity of the external persistent storage. Therefore, our
current implementation only analyzes queries that modify data (i.e., INSERT, UPDATE, and DELETE).
Since SELECT statements do not modify data, $P_v$ simply queues the \variants SELECT statements to
ensure the correct number of queries is received. No AST processing or string comparisons are
performed for SELECT statements. One exception is when an UPDATE or INSERT statement includes a
SELECT statement as a subquery. In this case, $P_v$ performs traversal and string comparison.

\begin{conference}
A limitation to this approach occurs when two servers insert different values (e.g., a timestamp)
  into the database but only a single value is truly stored. If the individual servers then perform
  a select on this value and compare the retrieved value to the original, at least one server will
  fail to pass this check. Fortunately, this case of extreme defensive programming was not
  encountered during testing with Mediawiki (Section~\ref{sec:eval}), and we leave addressing this
  issue for future work if there is a need to support this scenario.
\end{conference}

\section{Security Evaluation}
\label{sec:seceval}

\begin{conference}
  \begin{figure*}[h!]
  \centering
  \begin{subfigure}[b]{0.245\textwidth}
    \includegraphics[width=\textwidth]{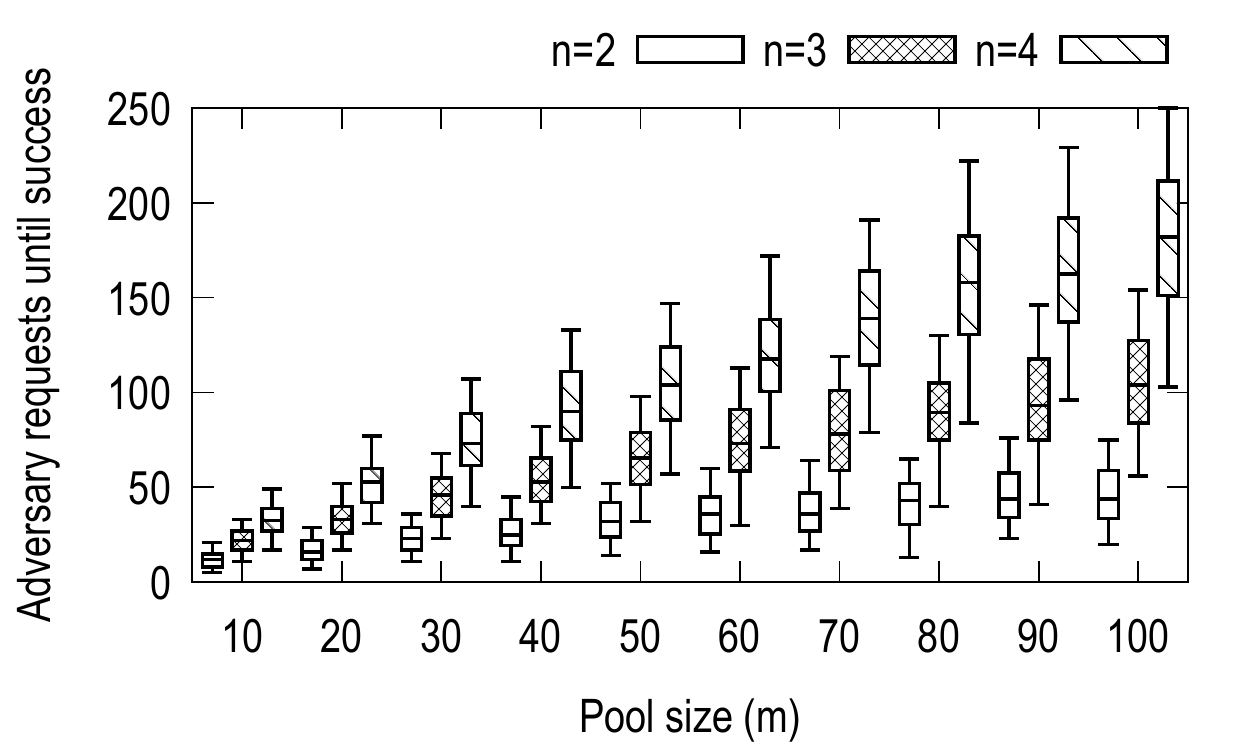}
    \caption{$\rejuvenatedPerAdvReq=1$}
    \label{fig:security:b=1}
  \end{subfigure}
  \begin{subfigure}[b]{0.245\textwidth}
    \includegraphics[width=\textwidth]{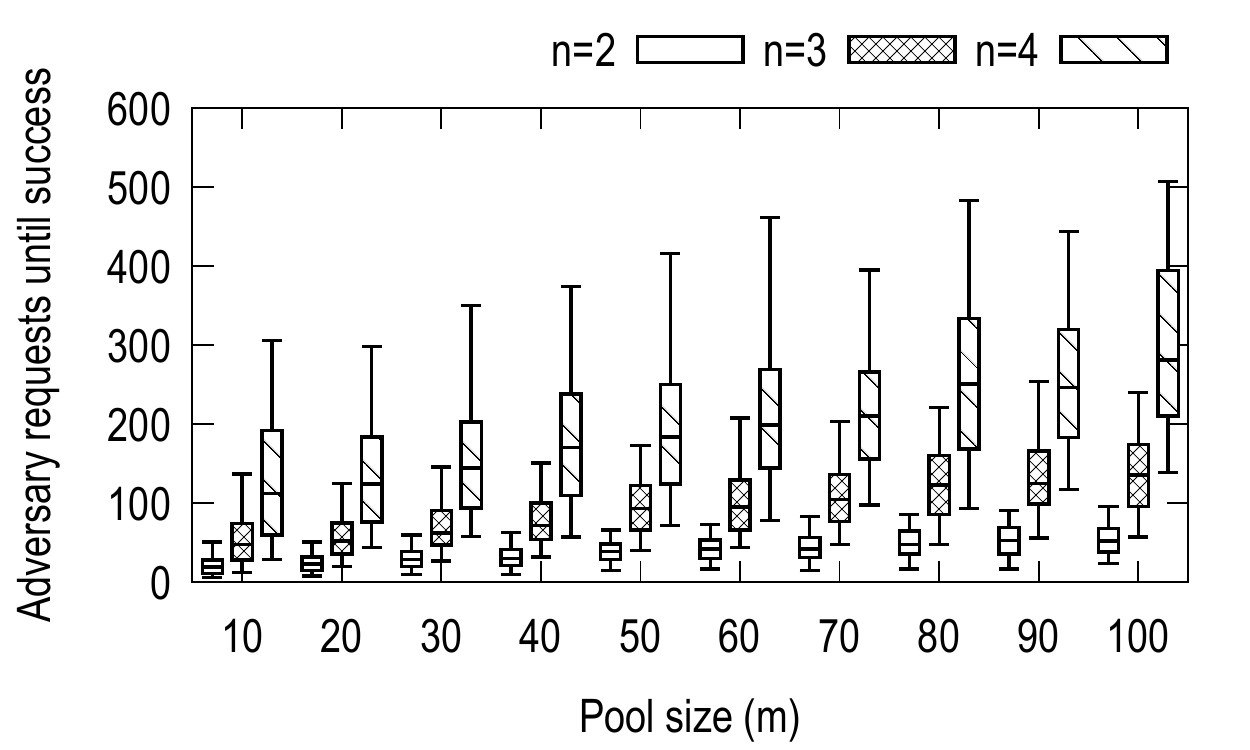}
    \caption{$\rejuvenatedPerAdvReq=3$}
    \label{fig:security:b=3}
  \end{subfigure}
  \begin{subfigure}[b]{0.245\textwidth}
    \includegraphics[width=\textwidth]{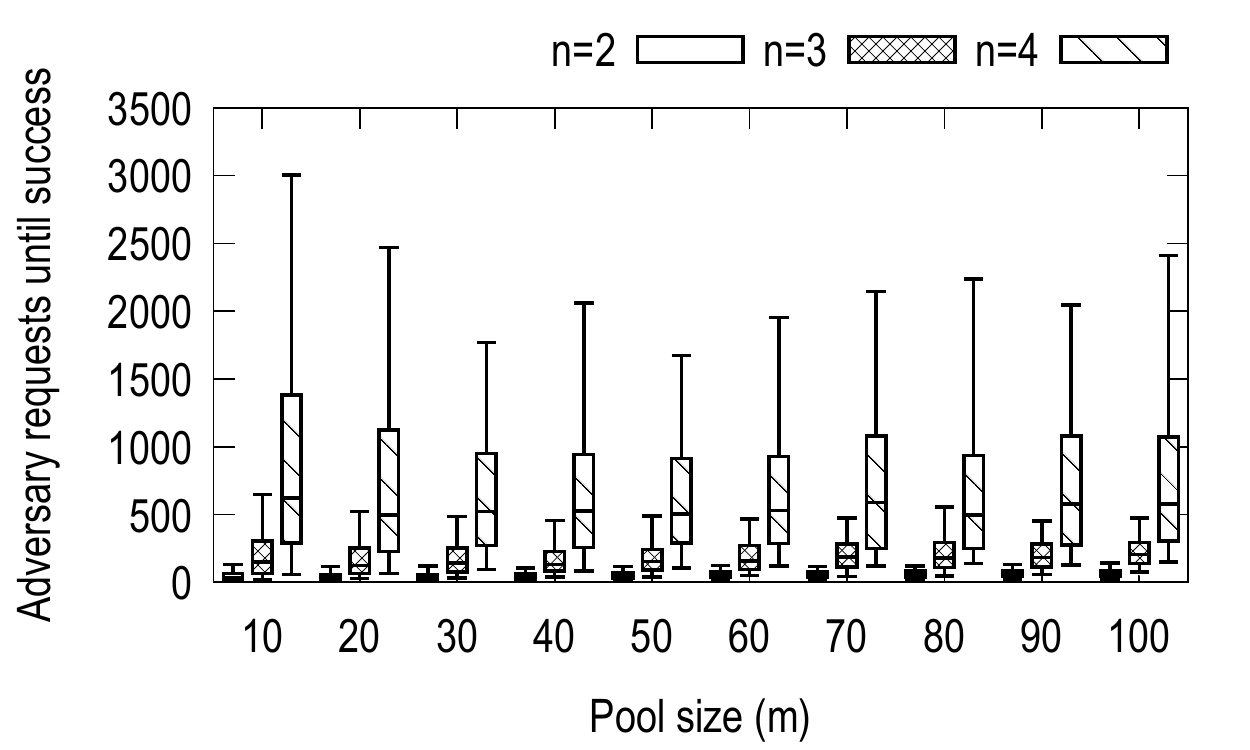}
    \caption{$\rejuvenatedPerAdvReq=5$}
    \label{fig:security:b=5}
  \end{subfigure}
    \begin{subfigure}[b]{0.245\textwidth}
      \includegraphics[width=\textwidth]{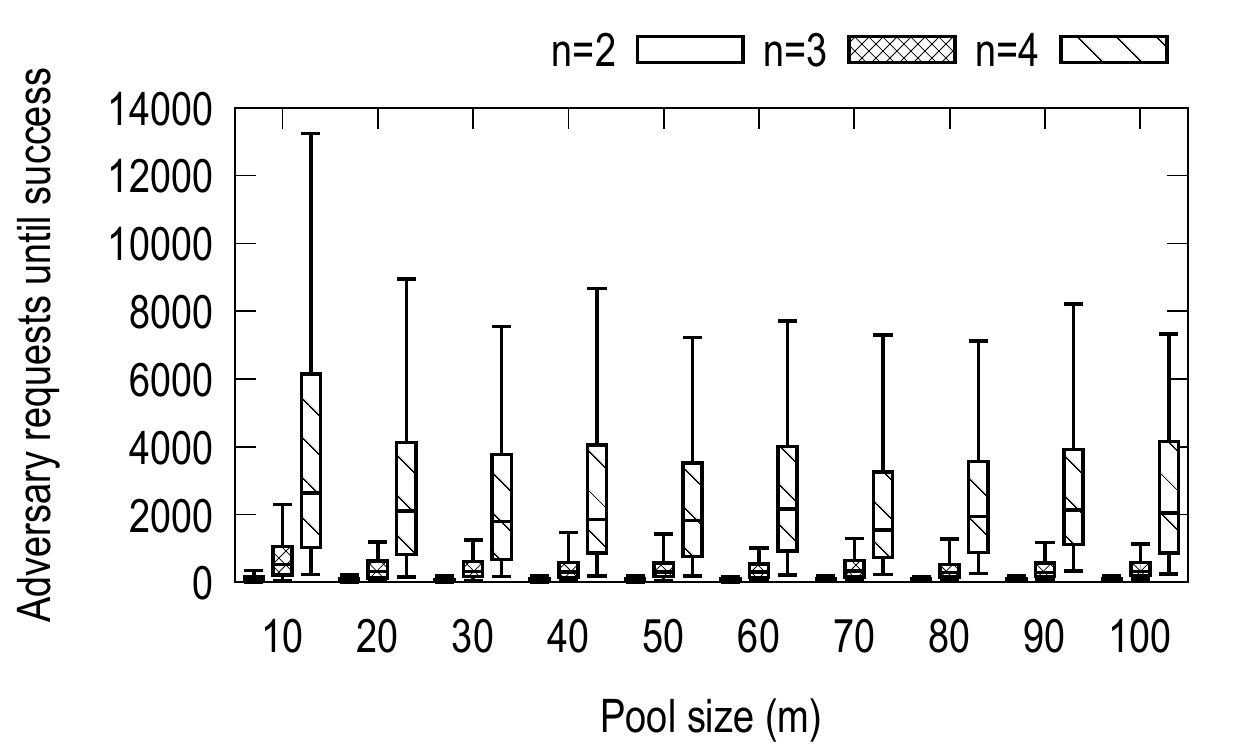}
    \caption{$\rejuvenatedPerAdvReq=7$}
    \label{fig:security:b=7}
  \end{subfigure}
  \caption{Distribution of \# adversary requests until success}
  \label{fig:security:b}
  \end{figure*}
\end{conference}

An adversary is successful when one of its HTTP requests is served by
a serving set containing already compromised replicas, since then all
of the corresponding SQL requests are controlled by the adversary.  In
this section, we analyze the probability that an adversary meets this
condition given a configuration of \variants variants, each
with \replicasPerVariant replicas.  We evaluate this probability in
two ways.  First, by modeling \systems using the well
known \emph{balanced allocations} problem, we argue that the fraction
of compromised replicas can never get very large.  Second, we evaluate
concrete configurations using simulations.

A \system randomly
refreshes on average \rejuvenatedPerReq server replicas per HTTP
request (Section~\ref{sec:replica_refresh}).  These replicas are selected from the entire
$\replicasPerVariant \variants$ replicas and not biased towards a
specific variant pool.  Our security evaluation is based
on \textit{adversarial} HTTP requests that compromise
replicas.  Therefore, we indirectly define \rejuvenatedPerReq using a
variable \rejuvenatedPerAdvReq, which defines the number of
replicas that are randomly refreshed between \textit{adversarial}
requests.  The number \rejuvenatedPerAdvReq of replica refreshes
per \textit{adversarial} request might not be immediately evident to
the defender.  This value can be estimated from the maximum
fraction \fractionAdvReqs of service requests that can be adversarial
and the number \rejuvenatedPerReq of server replica refreshes
performed per HTTP request, i.e., $\rejuvenatedPerAdvReq
= \ka$.  \fractionAdvReqs can, in
turn, be estimated from the expected overall HTTP request rate and the
expected adversarial HTTP request rate.

Of course, in practice there may be many benign HTTP requests that
occur between the adversary's HTTP requests that compromise a server
replica.  Furthermore, the adversary may require many HTTP requests to
compromise a server replica, e.g., due to memory defenses such as
ASLR.  Such defenses are further enhanced by the random serving set
selection created by \systems.  However, for simplicity, our
discussion assumes each adversarial HTTP request compromises a single
server replica.

Finally, our evaluation ignores the possibility that the
adversary can compromise both one uncompromised replica and, 
having a serving set of entirely corrupted replicas, then corrupt the
database, all in a \textit{single} HTTP request.  Allowing for this
possibility does not change our analysis qualitatively but complicates
our discussion.  As such, both our theoretical analysis and
our simulation results below assume that with a single adversary
request, the adversary can either corrupt one replica in its serving
set or, if that request is served by a serving set with all corrupt
replicas, can compromise the database.  It cannot do both with one
request, however.

\subsection{Theoretical Analysis}
\label{sec:analysis}

The adversary's goal is to be assigned a serving set only
containing already-compromised replicas.  Assume it has
compromised \compromised{\variantIdx} replicas of variant \variantIdx
and $\compromised{} = \sum_{\variantIdx =
1}^{\variants} \compromised{\variantIdx}$ replicas in total.  The
probability of selecting a compromised server in variant \variantIdx
is $\frac{\compromised{\variantIdx}}{\replicasPerVariant}$.  Therefore,
the probability of selecting a compromised server in all variants is
\[
\prob{\successEvent}
= \prod_{\variantIdx=1}^{\variants} \frac{\compromised{\variantIdx}}{\replicasPerVariant}
\le \left(\frac{(\compromised{}/\variants)}{\replicasPerVariant}\right)^{\variants}
\]
where the rightmost inequality follows because under the constraint
$\compromised{} = \sum_{\variantIdx =
1}^{\variants} \compromised{\variantIdx}$, the product
$\prod_{\variantIdx = 1}^{\variants} \compromised{\variantIdx}$ is
maximized when each $\compromised{\variantIdx}
= \compromised{}/\variants$.

Suppose that between serving adversary requests,
$\rejuvenatedPerAdvReq \ge 1$ replicas are chosen uniformly at random
from the $\variants\replicasPerVariant$ replicas and replaced with
rejuvenated versions.  Then, the number
\cleansedPerRequestRV{\compromised{}} of compromised replicas cleansed
after serving an adversary query after which there are \compromised{}
compromised replicas in total (and before serving the next adversary
query) is hypergeometrically distributed, i.e.,
$\cleansedPerRequestRV{\compromised{}} \sim
\hypergeometric{\compromised{}}{\variants\replicasPerVariant}{\rejuvenatedPerAdvReq}$.
Using a well-known tail-bound for the hypergeometric
distribution~\cite{chvatal:1979:tail},
\[
\prob{\cleansedPerRequestRV{\compromised{}} = 0} \le e^{-2\rejuvenatedPerAdvReq\left(\frac{\compromised{}}{\variants\replicasPerVariant}\right)^2} 
\]
So, if the adversary has compromised a fraction
\compromisedFraction of the replicas (i.e.,
$\frac{\compromised{}}{\variants\replicasPerVariant} =
\compromisedFraction < 1$), then at least one compromised replica is
cleansed with probability $\prob{\cleansedPerRequestRV{\compromised{}}
  \ge 1} \ge 1 - e^{-2\compromisedFraction^2\rejuvenatedPerAdvReq}$,
while $\prob{\successEvent} \le \compromisedFraction^{\variants}$.

We argue that \compromised{} and therefore \compromisedFraction will
tend to stay small, with high probability.  Specifically, cleansing
at least one compromised replica between adversary queries, with high
probability when \compromisedFraction is somewhat large, enables us to
leverage known results in \textit{balanced allocations} to reach this
conclusion.  In particular, Azar et al.~\cite{azar:1999:balanced}
considered an experiment in which a set of \balancedBinsSample bins is
chosen from among a total of \balancedBins bins uniformly at random,
and then a ball is placed into the least-full bin from among
these \balancedBinsSample.  After each such ball placement, a ball is
chosen uniformly from among all balls in the bins and removed, and
then the entire experiment (selection of
\balancedBinsSample bins, placement of a ball in the least full bin,
and then removal of a random ball) is repeated infinitely many times.
Azar et al.\ showed that in the stable distribution, the most-full bin
contains $\frac{\ln \ln \balancedBins}{\ln \balancedBinsSample} +
O(1)$ balls with high
probability~\cite[Theorem~1.2]{azar:1999:balanced}.  In our case, each
variant is analogous to a bin; each replica compromise is analogous to
a ball; the adversary is allowed to compromise any not-already-compromised
replica in its serving set of size \variants (i.e., $\balancedBinsSample =
\variants$); and each ball removal is analogous to a rejuvenation.
The most important difference between the problem considered by Azar
et al.\ and ours is that for Azar et al., the removal of a random ball
would, in our terminology, correspond to the certain rejuvenation of a
\textit{compromised} replica, i.e., 
$\prob{\cleansedPerRequestRV{\compromised{}} = 1} = 1$.  In our case,
$\prob{\cleansedPerRequestRV{\compromised{}} \ge 1} < 1$, but since
\prob{\cleansedPerRequestRV{\compromised{}} \ge 1} grows quickly with
\compromisedFraction, selecting an even modest \rejuvenatedPerAdvReq
is enough to ensure that \compromisedFraction tends to stay small.
As such, and because
$\frac{\ln \ln \balancedBins}{\ln \balancedBinsSample} = O(1)$ when
$\balancedBinsSample = \balancedBins$, we can expect that
$\compromised{\variantIdx} = O(1)$ with high probability.

\subsection{Simulation Analysis}
\label{sec:sim}

To more concretely illustrate the number \rejuvenatedPerAdvReq of
rejuvenations needed in various scenarios, we conducted a number of
simulations.  In these simulations, the adversary is presented a series
of requests to a (simulated) service, compromising the replica in each
request's serving set from the pool with the fewest compromised
replicas (and that was not already compromised).  The adversary did
this until it obtained a serving set of entirely compromised replicas.
Figure~\ref{fig:security:b} shows the distribution of the number of
\textit{adversary} queries until adversary success, where each boxplot
shows the first, second (median), and third quartiles, and whiskers
extend to the 5\textsuperscript{th} and 95\textsuperscript{th}
percentiles.  Each boxplot was computed from 200 trials.

Figure~\ref{fig:security:b} indicates that \variants and
\rejuvenatedPerAdvReq both have a substantial impact on security, as
also predicted above analytically, whereas the effect of
\replicasPerVariant is less pronounced.  The effect of increasing
\rejuvenatedPerAdvReq can be observed by noting the growth in the
y-axis as \rejuvenatedPerAdvReq is increased from
$\rejuvenatedPerAdvReq = 1$ in Figure~\ref{fig:security:b=1} through
$\rejuvenatedPerAdvReq = 7$ in Figure~\ref{fig:security:b=7}.  As
\rejuvenatedPerAdvReq is increased to a larger fraction of
\replicasPerVariant, the security improvement implied by increasing
\replicasPerVariant is muted; e.g., contrast the slope of the median
points for a given \variants when $\rejuvenatedPerAdvReq = 1$ in
Figure~\ref{fig:security:b=1} and those when $\rejuvenatedPerAdvReq =
7$ in Figure~\ref{fig:security:b=7}.

The main lesson from these simulations is that to maximize security,
it is most important to employ as many variants as possible (i.e.,
increase \variants) and to limit the fraction of requests that can be
adversarial (thereby increasing \rejuvenatedPerAdvReq).  The latter
might occur naturally, owing to a substantial legitimate load on the
service, or it might need to be imposed artificially, e.g., via
rate-limiting techniques akin to those used for defending against DoS
attacks.

\begin{table*}[ht]
  \centering
  \scriptsize
  \caption{Throughput and Latency comparison of \variants-25-Variant
  environments to the 30 VCPU baselines.}
  \label{tab:jmeter-overhead}
  \vspace{-1em}
\begin{tabular}{|l|l|l||l|l|}
\hline
\multicolumn{1}{|c|}{\textbf{}} & \multicolumn{2}{c||}{\textbf{Throughput}} & \multicolumn{2}{c|}{\textbf{Latency}} \\ \hline
\multicolumn{1}{|c|}{\textbf{Environment$^*$}} &
  \multicolumn{1}{c|}{\textbf{\begin{tabular}[c]{@{}c@{}}Caching Enabled
    Baseline\end{tabular}}} &
    \multicolumn{1}{c||}{\textbf{\begin{tabular}[c]{@{}c@{}}Caching
      Disabled Baseline\end{tabular}}} &
      \multicolumn{1}{c|}{\textbf{\begin{tabular}[c]{@{}c@{}}Caching
	Enabled Baseline\end{tabular}}} &
	\multicolumn{1}{c|}{\textbf{\begin{tabular}[c]{@{}c@{}}Caching
	Disabled Baseline\end{tabular}}} \\ \hline
2-25-Variant, \kn $ = 0$ & -55.95\% & -54.62\% & 307.86\% & 163.69\% \\ \hline
3-25-Variant, \kn $ = 0$ & -60.39\% & -59.20\% & 353.64\% & 193.29\% \\ \hline
4-25-Variant, \kn $ = 0$ & -69.86\% & -68.95\% & 497.01\% & 285.98\% \\ \hline
\multicolumn{5}{p{5.5in}}{$^*$Note the \system environments were
  configured to disable caching. The caching enabled and
  disabled refers to the baseline server configuration.}
\end{tabular}
\end{table*}

\section{Performance Evaluation}
\label{sec:eval}

In this section, we describe the prototype implementation, experimental setup, and conduct a
performance evaluation.

\subsection{Prototype Implementation}

Our prototype implements each component described in Section~\ref{sec:design}. The scheduling proxy
and verification proxy were both implemented in nearly 2,000 lines of Python code combined. The
query matching module was written in just under 200 lines of code, in addition to using a PostgreSQL
engine in C to parse the query. Finally, the Linux host agent was implemented in 600 lines of Python
code; the Windows host agent driver in 2,800 lines of C code; the DLL in 330 lines of C code; and
the IIS module in 40 lines of C\# code. The proxies and Linux hosts were implemented on Ubuntu
Server 16.04 (Xenial) cloud images, and the Windows hosts were implemented on Windows Server 2012
running IIS 8.0.

Our prototype simulates refreshing using a shim in the scheduling proxy, which makes servers appear
offline for 1~second after the refresh invocation. We simulated refreshing because the na\"ive
approach of reverting to an image snapshot in our OpenStack testbed took over 20 seconds. Amazon
Firecracker~\cite{firecracker} is a virtualization technology built on KVM allowing the deployment
of light-weight ``microVMs''. These microVMs provide the same isolation of traditional VMs but have
the performance of containers (i.e., launching a microVM in 125 ms). However, Firecracker was not
available at the time of implementation and its integration is left for future work. We believe our
simulated refresh is more than conservative. 

\subsection{Environment}
We hosted our test environment in CloudLab~\cite{cloudlab} using an OpenStack Queens bare metal
deployment with nine c6320 compute nodes (28 cores 2.00 GHz, 256 GB RAM, 10 Gb NIC) from the Clemson
cluster. One compute node was dedicated for each of: database server, scheduling proxy server,
verification proxy server, each variant pool, and clients generating the traffic.

The scheduling VM, verification proxy VM, and client VM were each assigned 56 VCPUs and 64 GB RAM.
Three baseline servers were assigned 50 VCPUS, 40 VCPUs, and 30 VCPUs each with 64 GB RAM. In
addition, each of the 100 variant web server VMs were assigned 2 cores and 4 GB RAM. The number of
VCPUs for each baseline server was chosen to reflect the number of total VCPUs in a single variant
pool for $\replicasPerVariant = 15, 20, 25$. The total number of VCPUs over all variant pools is a
resource cost associated with the desired \variants value and not considered when comparing to the
baseline performance. Our proxies require additional resources, but we did not consider finding an
optimal number of cores for each proxy that balances resource overhead with performance.

As previously discussed, we created two prototype host agents, one for an Apache Server running on
Ubuntu Server 16.04 and another for IIS 8.0 web server running on Windows Server 2012. However, for
ease of scaling our evaluation to four variants, our evaluation uses only the Apache/Ubuntu
environment to simulate four distinct variant environments. The Apache/Ubuntu host agent had better
supported tools for configuring the hosts and was easier to scale due to bugs in the Windows
implementation causing instability over long tests. Each server hosted Mediawiki version 1.18.6, an
older version that is compatible with a Wikipedia database dump from October 2007~\cite{trace}. Each
server was configured in the following ways: Mediawiki was configured to use a PostgreSQL backend;
PHP was configured to disable persistent database connections; the web servers were configured to
use CGI, disable all caching, and disable HTTP Keep Alive; and finally the Mediawiki application was
configured and modified to disable all caching in the variant servers. Unless otherwise stated above
or in the experiment descriptions below, all web server performance settings remained at their
default settings.

\begin{figure}[t]
  \centering
  \includegraphics[width=0.9\columnwidth,clip,trim=0.08in 0.00in 0.1in 0.1in]{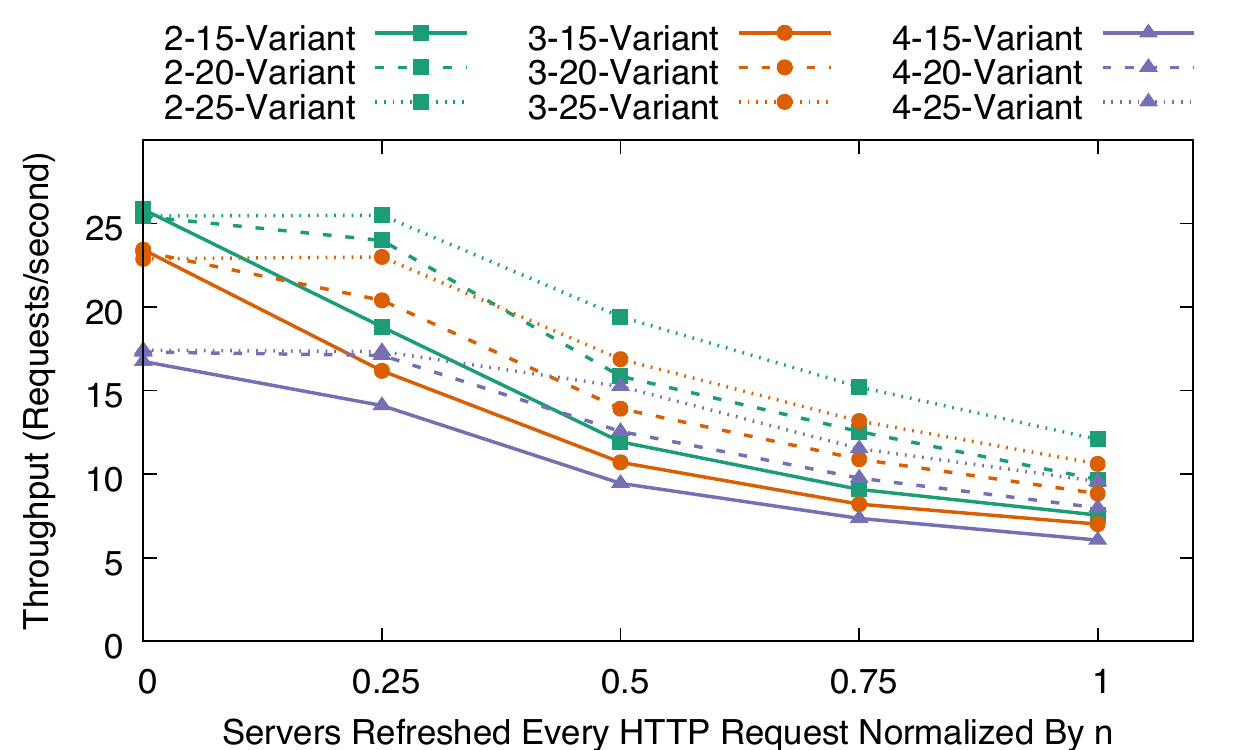}
  \caption{Average throughput serving the \textit{Protected\_area page} as \rejuvenatedPerReq is
  varied in each environment.}
  \label{fig:jmeter-throughput}
\end{figure}

\subsection{Performance Impact of \variants, \replicasPerVariant, and \rejuvenatedPerReq}
\label{sec:refreshing-eval}

The security evaluation in Section~\ref{sec:seceval} found that \variants and \rejuvenatedPerAdvReq
have substantial impacts on security. Since \rejuvenatedPerAdvReq is correlated to
\rejuvenatedPerReq, which determines how many servers are refreshed after each HTTP request,
increasing \rejuvenatedPerAdvReq to increase security also increases the number of refreshes after
each HTTP request. Increased refresh rates cause performance to suffer, since fewer servers will be
available to service incoming requests. This experiment explores how throughput and latency are
impacted as \rejuvenatedPerReq varies for different \variants, \replicasPerVariant configurations.

\subsubsection{Experimental Setup} 

\begin{figure}[t]
  \centering
  \includegraphics[width=0.9\columnwidth,clip,trim=0.08in 0.00in 0.1in 0.1in]{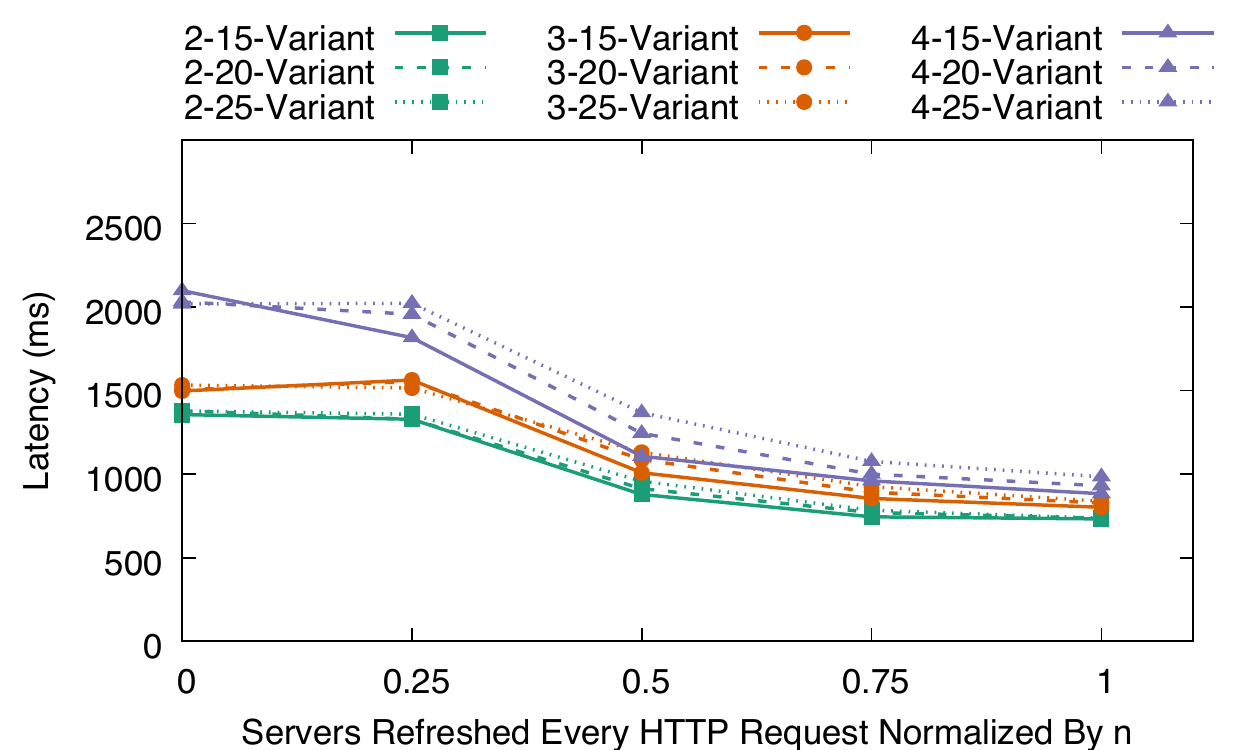}
  \caption{Average latency serving the \textit{Protected\_area} page as \rejuvenatedPerReq is varied in each environment.}
  \label{fig:jmeter-latency}
\end{figure}

Apache JMeter~\cite{jmeter} was configured to request a single web page from the Wikipedia snapshot
with 35 concurrent connections for 180 seconds. Since our prototype incurs overhead on correlating
and comparing queries at the verification proxy, pages that require more queries to load result in
larger overheads. For this test, the web page was chosen so the number of queries between the server
and the database needed to generate the web page
results in the mean number of queries, among all hosted pages. This was determined by requesting
every page hosted by the application and recording the number of queries from the web server to the
database needed to generate the page. As a result the \textit{Protected\_area} page was chosen for
this experiment and each JMeter worker was configured to request this page.

Using the above JMeter configuration, for each chosen \variants and \replicasPerVariant
configuration we ran a test increasing \rejuvenatedPerReq normalized by \variants (e.g., \kn) from
$0$ to $1.0$ in $0.25$ increments. The JMeter test also ran for 30 VCPU, 40 VCPU, and 50 VCPU
baseline servers with caching enabled and caching disabled. For each test, we recorded the number of
successful (HTTP 200) responses per second handled by the web environment and each response's
latency.

\begin{conference}
  Our analysis only considers the HTTP 200 responses (e.g., goodput). This is due to configurations
  with higher unavailability servicing magnitudes more total requests during the experiment and a
  fraction of those requests receive HTTP 200 responses. If the analysis considered the HTTP 503
  responses, configurations that produce more HTTP 503 responses would falsely indicate higher
  throughput and lower latency. Therefore, we consider our analysis to be conservative and justly
  the represent the overhead.
\end{conference}

\subsubsection{Throughput Results}
\label{sec:throughputresults}

Our first analysis considers the throughput of the baseline servers and various \variants,
\replicasPerVariant, and \rejuvenatedPerReq configurations. We observed the throughput for each of
the caching and non-caching baseline environments was barely impacted by the 30, 40, and 50 VCPU
configurations. To conservatively report our overhead, we only compare our environments to the 30
VCPU caching and non-caching baseline, which observed 57.794 and 56.111 requests per second
respectively.
We note the result of adding VCPUs to the baseline servers and observing no impact on the throughput
is potentially a result of hosting dynamic web pages that require many SQL queries over the network
to a remote database. 

Table~\ref{tab:jmeter-overhead} shows the impact of our \system prototype without refreshing ($\kn =
0$). For brevity, we will only discuss the overhead with respect to the cached baseline as there is
not a substantial difference in throughput between the two baseline environments.  For the
$\replicasPerVariant = 25$ environments, we observed a 56\% reduction (-32.33 RPS) in throughput for
the 2-25 environment and up to a 70\% reduction (-40.37 RPS) in throughput for the 4-25 environment.
Figure~\ref{fig:jmeter-throughput} depicts the impact of increasing \rejuvenatedPerReq, which
reveals the following three trends:

\myparagraph{Increasing \variants reduces throughput}
As \variants is increased, throughput decreases, including a 5 request-per-second drop from
$\variants = 3$ to $\variants = 4$. This is due to two reasons. First, with more variants, the
verification proxy needs to queue and correlate connections from more servers in a serving set.
Second, the overhead of parsing and comparing queries at the verification proxy linearly grows with
\variants. Note this overhead is negligible on a single query, however since pages require many
queries to generate content, the overhead compounds and results in noticeable performance
degradation.

\myparagraph{Increasing \rejuvenatedPerReq reduces throughput}
For each \variants-\replicasPerVariant configuration, lower \rejuvenatedPerReq  values result in a
higher number of successful requests per second. As \rejuvenatedPerReq is increased and replicas go
offline more frequently, fewer requests can be handled. Once $\kn \geq 1$ the performance converges
for each environment. In fact, when $\kn = 1$ we have a situation where each server in a serving set
is refreshed after it is used to service a request, and thus it is meaningless to have
$\rejuvenatedPerReq > \variants$. This scenario behaves similar to a system where a serving set is
spawned on demand to service every web request. Notably for \kn = 0.5 we observe a minimum
throughput reduction of 65\% for $\variants = 2, \replicasPerVariant = 25$ and a maximum throughput
reduction of 83\% for $\variants = 4, \replicasPerVariant = 15$. Of further interest is for $\kn =
.25$ we observe no drop in throughput for large enough values of \replicasPerVariant, i.e.,
$\replicasPerVariant = 25$. 

\myparagraph{Increasing \replicasPerVariant allows higher throughput for higher \rejuvenatedPerReq}
Finally, increasing \replicasPerVariant allows an environment to provide a higher throughput for a
given \variants while increasing \rejuvenatedPerReq. This is due to the larger number of possible
serving sets that are available to service incoming requests.

\subsubsection{Latency Results}

Our second analysis considers the latency of the baselines and various \variants,
\replicasPerVariant, and \rejuvenatedPerReq configurations. Similar to the comparison to the
baseline in the throughput analysis, we observed minimal differences in latency for the different
baseline VCPU configurations and compare the \system environments to the caching and non-caching 30
VCPU baselines which observed a 338.19~ms and 523.09~ms latency respectively. 

Table~\ref{tab:jmeter-overhead} shows the overhead of our \system prototype without refreshing ($\kn
= 0$) comparing against caching enabled and disabled baselines. For the $\replicasPerVariant = 25$
environments and caching enabled, we observed a 308\% increase (1041~ms) in latency for $\variants =
2$ and up to a 497\% increase (1681~ms) for $\variants  = 4$. However, a fairer comparison is to the
caching disabled baseline, since each variant disabled caching. Comparing to this baseline we see
164\% increase (856~ms) in latency for $\variants  = 2$ and up to a 286\% (1496~ms) increase for
$\variants  = 4$. Figure~\ref{fig:jmeter-latency} depicts the impact of increasing
\rejuvenatedPerReq, which reveals the following two trends:

\myparagraph{Increasing \variants increases latency}
Similar to the throughput results, we see \variants has an impact on latency, with a 500~ms jump
occurring from $\variants = 3$ to $\variants = 4$. This is caused by negligible query processing
overhead compounding at the verification proxy as \variants is increased. However, unlike in the
throughput results, we do not observe a major impact on latency by increasing \replicasPerVariant
for any given \variants, but smaller \replicasPerVariant, which result in lower throughput, provide
incremental latency improvements. Notably, for $\kn = 0.5$ we observe a 83\% (435~ms) increase in
latency for the $\variants = 2, \replicasPerVariant = 25$ environment and a 111\% (584~ms) increase
in latency for the $\variants = 4, \replicasPerVariant = 15$ environment.

\myparagraph{Increasing \rejuvenatedPerReq decreases latency}
As \rejuvenatedPerReq is increased and throughput is decreased (as discussed
in~\ref{sec:throughputresults}), we observed a decrease of latency in each environment. For example,
for $\replicasPerVariant  = 25$ environments when $\kn = 1$, the overhead compared to the baseline
is much lower than compared to $\kn = 0$. When compared to the baseline with caching enabled, we
observed a 118\% increase (398~ms) in latency for the $\variants  = 2$ environment and up to a 191\%
(647~ms) increase in latency for the $\variants  = 4$ environment. Further, when compared to the
baselines with caching disabled, we observed a 41\% increase (213~ms) in latency for $\variants  =
2$ and up to a 88\% increase (462~ms) for $\variants  = 4$.

These trends can be explained by the number of concurrent requests handled by the environment. For
example, when \rejuvenatedPerReq is low and the throughput is high, the proxies are contending for
resources to process all the independent requests.  However, as we increase \rejuvenatedPerReq and
decrease the throughput, the verification proxy is able to dedicate more resources to quickly
comparing the queries, which results in quicker responses to clients.

\subsection{Tuning Security and Cost}
\label{sec:eval_cost}

\systems assume a powerful adversary with exploits for all \variants variants and can perform those
exploits with a single request (e.g., ASLR may prevent the latter). We also assume the adversary is
fully aware of the \system defense mechanism and can strategically form an optimal attack plan
(Section~\ref{sec:seceval}). In this section, we describe how \variants, \replicasPerVariant, and
\rejuvenatedPerReq impact the period of resistance and monetary cost of additional VCPUs.

Note that practical cost prevents \systems from providing resistance to this powerful adversary in
perpetuity. Rather, \system provides an invaluable delay that allows offline IDS (manually confirmed
or otherwise) to catch up. Furthermore, less powerful adversaries are even less likely to succeed.

\subsubsection{Period of Resistance}

\begin{table}[t]
\centering
\scriptsize
  \caption{Equations of trend lines.}
  \label{tab:equations}
  \vspace{-1em}
\begin{tabular}{|c|c|c|c|}
\hline
  & \textbf{\variants = 2} & \textbf{\variants = 3} & \textbf{\variants = 4} \\
\hline
  \textbf{Equation} & $y = 14.235e^{0.2028\rejuvenatedPerAdvReq}$ &
  $y = 25.46e^{0.3446\rejuvenatedPerAdvReq}$ &
  $y = 41.537e^{0.5057\rejuvenatedPerAdvReq}$  \\
\hline
  \textbf{$R^2$} & 0.9923 & 0.9903 & 0.9797 \\
\hline
\end{tabular}
\end{table}

\begin{table}
\centering
\scriptsize
  \caption{Size of windows, expressed in time, for \fractionAdvReqs = 10\%.}
  \label{tab:windows}
  \vspace{-1em}
\begin{tabular}{|c|c|c|c|} 
\hline
\textbf{\variants} & \textbf{\kn = .25} & \textbf{\kn =.5} & \textbf{\kn = .75}  \\ 
\hline
\textbf{2} & 39 sec. & 108 sec. & 298 sec. \\ 
\hline
\textbf{3} & 337 sec. & 74 min. & 16.4 hours \\ 
\hline
\textbf{4} & 108 min. & 11.8 days & 5.1 years\\
\hline
\end{tabular}
\end{table}

To bridge the theoretical analysis in Section~\ref{sec:seceval} and the performance analysis earlier
in Section~\ref{sec:eval}, we selected several configurations for comparison.
Specifically, we consider the three $\replicasPerVariant = 25$ environments as they provided the
best throughput for their respective \variants values.  Using $\variants = 2, 3, 4$ and
$\replicasPerVariant = 25$, we ran simulations from Section~\ref{sec:seceval} for
$\rejuvenatedPerAdvReq = 1 ... 15$. We then did a statistical analysis of the median
requests-until-compromise for each $\replicasPerVariant=25$ configuration, varying
\rejuvenatedPerAdvReq to fit a line to the curve.

Table~\ref{tab:equations} shows the resultant equations of the trend lines fit to each
$\replicasPerVariant=25$ configuration. All trends reported a $R^2$ value over 0.979. Using these
equations, we then plot the number of requests until adversary success, with respect to the percent
of malicious requests (\fractionAdvReqs), for each $\replicasPerVariant=25$ configuration and $\kn =
0.25, 0.5, 0.75$. Note the equations in Table~\ref{tab:equations} are in terms of
\rejuvenatedPerAdvReq, the number of servers refreshed between adversary requests, and so we use the
equality $\rejuvenatedPerAdvReq  = \ka$ to appropriately generate this graph.

\begin{figure}[t]
  \centering
  \includegraphics[width=0.9\columnwidth,clip,trim=0.08in 0.00in 0.1in 0.1in]{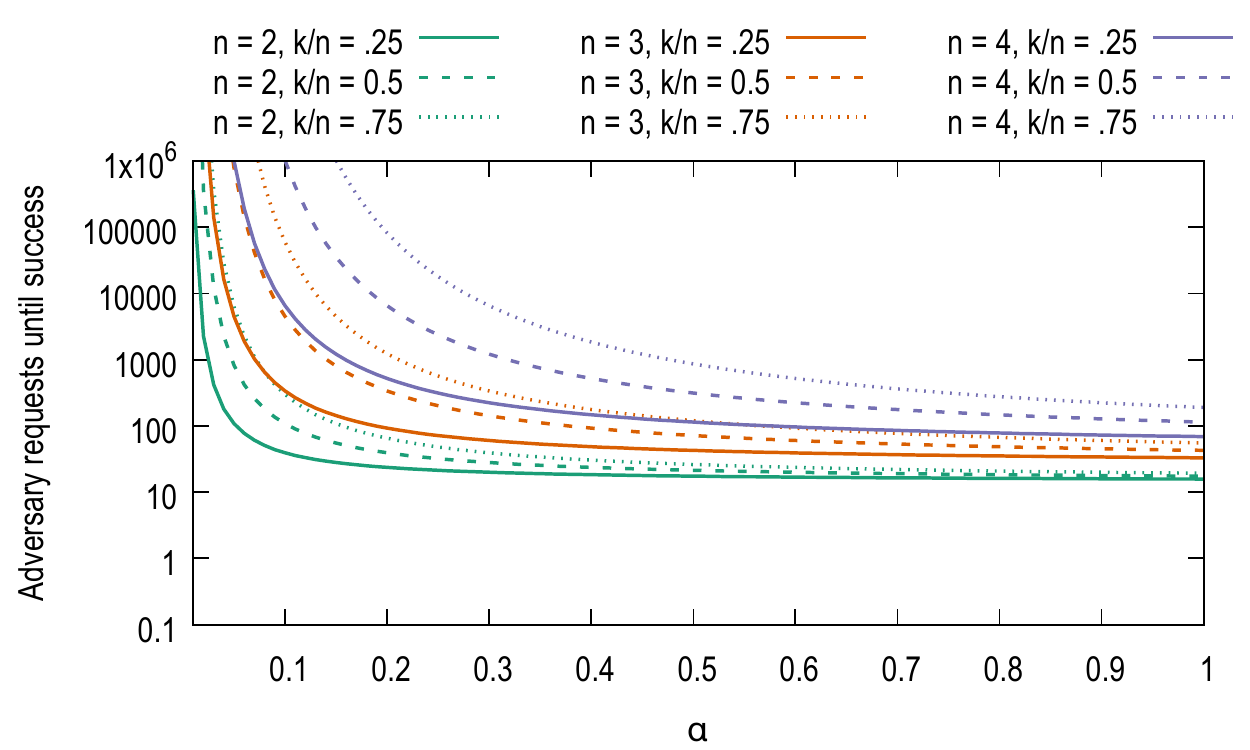}
  \caption{Number of requests until adversary success.}
  \label{fig:costeval}
\end{figure}

Figure~\ref{fig:costeval} shows the number of expected adversarial requests until success for each
$\replicasPerVariant = 25$ configuration and $\kn = .25$, $.5$, and $.75$. To reiterate our
takeaways from Section~\ref{sec:seceval}: (1) \variants and \rejuvenatedPerAdvReq (\ka) have the
largest impact on the security (2) \replicasPerVariant has a smaller impact on the system's security
guarantees (3) if $\rejuvenatedPerReq < \variants$, the adversary will eventually succeed. However,
different configurations provide larger periods of resistance  where an adversary will not succeed.
Next we will walk through a simple example to illustrate this point.

Consider a request rate of 10 requests per second and $\fractionAdvReqs = 0.1$ (e.g., adversary
request rate of 1 request per second). We assume both benign and malicious traffic rates are
constant to simplify the example. Using these values we now define a period of resistance, in
seconds, by using the y-axis of Figure~\ref{fig:costeval}, since the adversary sends 1 request every
second and the y-axis represents the median number of requests until success. Results are summarized
in Table~\ref{tab:windows}.

First consider the $\variants = 2$. For $\kn = .25$ the adversary is expected to succeed in under 40
seconds; for $\kn = .5$ this time is increased to 108 seconds; and finally for $\kn = .75$ it is
increased to just under 5 minutes. Now consider $\variants = 3$: for $\kn = .25$ the adversary
expects to succeed in just over 5 minutes; for $\kn = .5$, it succeeds in about 75 minutes; and
finally for $\kn = .75$ the adversary takes almost 16.5 hours to succeed. Finally, for $\variants =
4$ this trend continues.  For $\kn = .25$ the adversary expects to succeed in 108 minutes, followed
by just under 12 days for $\kn = .5$, and finally for $\kn = .75$ the adversary is expected to
attack without success for over 5 \textit{years}.

Through this trivial illustration, it is clear to see how much of a security impact using high
\variants and \rejuvenatedPerReq has on the environment; however, this also has the highest impact
on performance. Thus to complete this analysis and allow system administrators to determine which
parameters work best for their environment, we need to perform a cost analysis of \systems.

\subsubsection{Resource Cost}

The security of different environments comes with a resource cost. For each of the \variants and
\replicasPerVariant configurations we analyze the increase of VCPUs from the baseline and calculate
a value of Throughput per VCPU (\throughputPerVCPU), which visualizes the cost of resource
duplication as a return on investment (ROI). The increase of VCPUs in each environment trivially
allows an administrator to calculate the monetary cost overhead of implementing each configuration
by using current cloud computing costs.

Table~\ref{tab:resource-overhead} summarizes the results of this analysis for each of the \variants,
\replicasPerVariant, and $\kn = 0.5$ configurations from Section~\ref{sec:throughputresults}. Note
the first row represents the 30 VCPU non-caching baseline and we included the resistance of each
environment using Table~\ref{tab:windows}. As expected, the greater the \variants value the less
contribution any VCPU has on the overall system throughput. However, this results in a much greater
contribution to the period of resistance. Conversely, by increasing \replicasPerVariant, the
throughput of a given \variants-\replicasPerVariant-Variant environment can be regained with a small
decrease in the ROI.

%

\begin{table}[t]
  \centering
  \scriptsize
  \caption{Cost of \system environments compared to the 30 VCPU
  non-caching baseline (Section~\ref{sec:throughputresults}).}
  \label{tab:resource-overhead}
  \vspace{-1em}
\begin{tabular}{|l|l|l|l|l|l|}
\hline
\multicolumn{1}{|c|}{\textbf{n}} & \multicolumn{1}{c|}{\textbf{m}} &
  \multicolumn{1}{c|}{\textbf{\begin{tabular}[c]{@{}c@{}}VCPU Increase\end{tabular}}} &
    \multicolumn{1}{c|}{\textbf{\makecell{Throughput\\(RPS)}}} &
    \multicolumn{1}{c|}{\textbf{Throughput/VCPUs}} &
    \multicolumn{1}{c|}{\textbf{Resistance$^*$}} \\ \hline \hline
  - & -  & -        & 56.11 & 1.87 & 0     \\ \hline \hline
  2 & 15 & 100\%    & 11.94 & .199 & 108 sec.\\ \hline
  2 & 20 & 166.67\% & 15.87 & .198 & 108 sec. \\ \hline
  2 & 25 & 233.33\% & 19.43 & .194 & 108 sec. \\ \hline \hline
  3 & 15 & 200\%    & 10.71 & .119 & 74 min. \\ \hline
  3 & 20 & 300\%    & 13.93 & .116 & 74 min. \\ \hline
  3 & 25 & 400\%    & 16.88 & .112 & 74 min. \\ \hline \hline
  4 & 15 & 300\%    & 9.46  & .079 & 11.8 days \\ \hline
  4 & 20 & 433.33\% & 12.56 & .079 & 11.8 days \\ \hline
  4 & 25 & 566.67\% & 15.26 & .076 & 11.8 days \\ \hline
  \multicolumn{6}{l}{$^*$Median time for $\kn = 0.5$, $\fractionAdvReqs =
  0.1$.}
\end{tabular}
\end{table}

\section{Discussion}
\label{sec:discussion}

\myparagraph{Denial-of-Service}
Any solution reducing throughput makes denial-of-service attacks
easier. This is true in the case of \systems, which return a 503
Service Unavailable response if there are no open serving sets. Although
we enhance the availability of prior BFT and MVEE works, DoS is still a
threat. Note, traditional DoS mitigation can complement \systems by
alleviating the impact of attacks. However, \systems still provide value to
servers with lower request rates, such as an internal server opposed
to a popular publicly accessible web server. In the scenario of an internal
server, the adversary is assumed to have compromised a machine on the network
and makes unprivileged requests to the server from local network. DoS
attacks launched from the local network may be easier for admins to identify and stop.

\myparagraph{Application Determinism}
Our system cannot handle applications with specific types of
non-determinism, which differ from non-deterministic database fields
already discussed. For example, the ``Random Page'' link in Mediawiki
loads a random page from the server. As the individual replicas select
a random page, the queries will diverge on accesses to different pages. Such
non-deterministic features are currently not supported by \systems. Note
prior MVEE works (e.g., Orchestra~\cite{sjg+09}) also encounter this issue,
but solved it by intercepting system calls that have non-deterministic
output (e.g., \texttt{getrandom}), then make the call once
and copy the result to each variant. However, these prior works had
the advantage of residing on a single host, as such we leave the design and
integration of such a mechanism for future work.

\myparagraph{Period of Security Resistance}
\systems allows an administrator to balance security, resource
allocation, and performance.  If the administrator assumes the most powerful
adversary with exploits for all \variants variants,
Section~\ref{sec:eval_cost} demonstrated practical configurations that provide
a period of resistance ranging from seconds, to minutes, to days.  With
this strong threat model, even a short period of resistance can provide
invaluable defense.  For example, it can provide time for an offline
intrusion-detection system (IDS) to determine that an attempt to compromise
replicas is underway, before those compromises can result in a corruption of
the persistent storage.  Furthermore, since statistical IDS can raise false
alarms~\cite{axelsson:2000:base-rate}, the period can also give time for human investigation.

\section{Conclusion}
\label{sec:conc}

This work introduced \systems, an adversarial-resistant software
rejuvenation framework for cloud-based web applications. We improved
state-of-the-art intrusion-tolerant frameworks with the introduction of
the variable \replicasPerVariant, which increases the availability of
these systems and allows administrators to tune their environments to
balance resource and performance overhead with security guarantees
obtained. Through theoretical analysis and a performance evaluation of
our prototype, this work demonstrated the practicality of the \systems
framework.

\section*{Acknowledgements}
This work was supported by the National Science Foundation (NSF) SaTC grants CNS-1330553 and CNS-1330599. Opinions, findings, conclusions, or recommendations in this work are those of the authors and do not reflect the views of the funders.

\bibliographystyle{ACM-Reference-Format}
\bibliography{bib/ms}

\end{document}